\documentclass[usenatbib]{mn2e}

\usepackage{graphicx}
\usepackage{color}
\usepackage[space]{grffile}
\usepackage{latexsym}
\usepackage{textcomp}
\usepackage{longtable}
\usepackage{multirow,booktabs}
\usepackage{amsfonts,amsmath,amssymb}
\usepackage{url}
\usepackage{hyperref}
\usepackage[english]{babel}
\usepackage{graphicx}
\usepackage[space]{grffile}
\usepackage{url}
\usepackage[utf8]{inputenc}
\usepackage{hyperref}
\usepackage{longtable}
\usepackage{natbib}

\hypersetup{colorlinks=false,pdfborder={0 0 0}}

\title[Going, going, gone dark: Quantifying the scatter in the faintest dwarf galaxies]{Going, going, gone dark: Quantifying the Scatter in the Faintest Dwarf Galaxies}

\author[Ferah Munshi et al.]{Ferah Munshi$^{1,2}$, Alyson M. Brooks$^{1}$, Elaad Applebaum$^{1}$, Daniel R. Weisz$^{3}$, \newauthor Fabio Governato$^{4}$,  Thomas R. Quinn$^{4}$%, Sydney Duncan$^{4}$
\vspace*{6pt} \\
$^{1}$Department of Physics \& Astronomy, Rutgers, The State University of New Jersey, 136 Frelinghuysen Rd, Piscataway, NJ 08854;\\
$^{2}$Department of Physics and Astronomy, Vanderbilt University, PMB 401807, Nashville, TN 37206;\\
$^{3}$Department of Astronomy, 501 Campbell Hall \#3411, University of California at Berkeley, Berkeley, CA 94720-3411\\
$^{4}$Department of Astronomy, University of Washington, Box 351580, Seattle, WA 98195-1580; \\
}

\begin{document}

\maketitle

\label{firstpage}

\begin{abstract}

We predict the stellar mass-halo mass (SMHM) relationship for dwarf galaxies and their satellites residing in halos down to M$_{halo} =$ 10$^7$ M$_{\odot}$ with 10$^4$ M$_{\odot} <$ M$_{star}$($z=0$) $< 10^8$ M$_{\odot}$, and quantify the predicted scatter in the relation at the low mass end, using cosmological simulations.
%We examine the SMHM relationship for simulated dwarf galaxies with 10$^4$ M$_{\odot} <$ M$_{star}$($z=0$) $< 10^8$ M$_{\odot}$.  
The galaxies were drawn from a cosmological simulation %of a cosmological ``sheet'' of dwarf galaxies, 
run with the N-body + SPH code, ChaNGA, at a high resolution of 60 pc.
For M$_{halo} > 10^9$ M$_{\odot}$, the simulated SMHM relationship agrees with literature determinations, %derived from abundance matching results at high masses, 
including exhibiting a small scatter.   However, the scatter in the SMHM relation increases dramatically for lower-mass halos.  
We find that some of this scatter is due to {\em dark dwarfs}, halos devoid of stars. However, even when only considering well-resolved halos that contain a stellar population, the scatter in stellar mass reaches nearly 1 dex for M$_{halo}$($z=0$) 10$^7$ M$_{\odot}$. Much of this scatter is due to including satellites of the dwarf galaxies that have had their halo masses reduced through tidal stripping.
%When considering only luminous, well-resolved simulated galaxies, the scatter in stellar mass reaches nearly 1 dex for M$_{halo}$($z=0$) 10$^7$ M$_{\odot}$.  {\bf Much of this scatter is due to the inclusion of satellite galaxies } 
The fraction of dark dwarfs (those that contain no stars) increases substantially with decreasing halo mass. %For M$_{halo}$($z=0$) $< 10^8$ M$_{\odot}$, 50-80\% of halos contain no stars.
When these dark halos are considered, the true scatter in the SMHM at low masses is even larger. %: 2.4 dex at 10$^7$ M$_{\odot}$.  
%In summary, the assumption of a monotonic relationship between stellar mass and halo mass breaks down at low masses, implying that simple abundance matching prescriptions should be used with extreme caution.   
At the faintest end of the SMHM relation probed by our simulations, a galaxy cannot be assigned a unique halo mass based solely on its luminosity.  We provide a formula to stochastically populate low-mass halos following our results. %Finally, we show that the scatter we predict has implications for the the smallest masses in the stellar mass function, in the mass range currently probed by the newest DES dwarf galaxies.  
Our predicted large scatter at low halo masses increases the slope of the resulting stellar mass function on the ultra-faint dwarf galaxy scales currently being probed by such surveys as the Dark Energy Survey or the Hyper-Suprime Cam Subaru Strategic Program, and in the future by the Large Synoptic Survey Telescope.

\end{abstract}%

\begin{keywords}
galaxies: dwarf -- galaxies: haloes -- galaxies: formation 
\end{keywords}

\section{Introduction}
  In the $\Lambda$ Cold Dark Matter ($\Lambda$CDM) paradigm of cosmological structure formation, dwarf galaxies are predicted to be the smallest, most abundant, yet least luminous systems in the Universe. %In this paradigm, dwarfs are thought to have collapsed early to become the building blocks for larger halos and contributed significantly to the reionization of the universe. 
Recent attempts to link dwarf galaxies to their parent dark matter halos via abundance matching have led to discrepancies between theory and observations \citep[e.g.,][]{Ferrero2012,GK2014b, Papastergis2015, BDC2015}.  %Generally regarded as assumption free, 
Abundance matching quite literally matches a stellar mass or luminosity at a given abundance to dark matter halos with the same abundance,  derived from a dark matter-only simulation.  A monotonic relationship is generally assumed \citep{guo10,Behroozi2013,Moster2013}.  
For halos of roughly Milky Way mass and greater, abundance matching also reproduces clustering statistics \citep[e.g.,][and references within]{ConroyWechsler2009}.  Additionally, numerous abundance matching studies yield fair agreement for the stellar mass-to-halo mass relation (SMHM) for halos of masses $\gtrsim 10^{11}$ M$_{\odot}$. 

However, derivations of the SMHM at lower masses have yielded discrepancies \citep[e.g.,][]{Moster2013, Behroozi2013, Brook2014, GK2014}. If the SMHM has the form M$_{star} \propto$ M$^{\alpha}_{halo}$, the range of derived $\alpha$ varies from 1.6 -- 3.1 for low mass galaxies.
%\textcolor{red}{Discuss the range.  Moster, Behroozi only derived down to x,y halo masses, respectively.  And extrapolation from their fit at lowest masses suggests a,b slopes.  Brook, G-K tried to fit to lower masses using local group data.  Found 1.9 or 3.1.}  
\citet{Moster2013} and \citet{Behroozi2013} did not have data in order to derive the SMHM relation below stellar masses of a few $\times$10$^{7}$ M$_{\odot}$.  While their results from higher masses should obviously not be extrapolated to lower masses, the slopes at their lowest measured  mass were quite different, $\alpha = 1.4$ \citep{Behroozi2013} versus $\alpha = 2.4$ \citep{Moster2013}.  \citet{Brook2014} and \citet{GK2014}, on the other hand, used Local Group galaxy data to determine the SMHM relation at 10$^6 < $M$_{star}$/M$_{\odot} < 10^8$.  Again, they came to quite different conclusions about the value of $\alpha$, 3.1 in \citet{Brook2014} and 1.9 in \citet{GK2014}.\footnote{Though the slope is dependent on the normalization at higher masses, and the values come into better agreement when a consistent normalization is adopted, see \citet{GK2016}.} \citet{Read2017} find a shallower stellar mass function below $10^{9} M_{\odot}$, $\alpha = 1.6$, and attribute this to that fact that they use a strictly isolated galaxy sample, arguing that including galaxies processed in a group environment leads to a steeper SMHM relation.

%\textcolor{red}{High mass end small scatter . used baryonic simulations to point out the importance of dark halos.  Found bend. \citet{GK2016} recent paper on importance of scatter.  These aren't necessarily same thing...}
The relative agreement in the SMHM relation at the high mass end also points to the fact that the scatter at the high mass end is consistently measured to be relatively small, 0.2 dex or less \citep{Behroozi2013, Reddick2013, Kravtsov2014, Eagle2017}.  However, the scatter at the low mass end may be much larger.  In fact, the star formation history at dwarf galaxy scales is likely to depend on the mass accretion history of the halo \citep{BZ2014, Weisz2015a, Sawala2016}, and observations indicate that dwarf galaxies over a range of stellar masses may all occupy dark matter halos with a narrow range of mass \citep[e.g., see Figure 1 of][]{Klypin2015, Ferrero2012, strigari08}.  \citet{GK2016} explored the implications of scatter within M$_{star}$ at a given M$_{halo}$ using Local Group data down to M$_{star} \sim 10^5$ M$_{\odot}$.  They demonstrated that there is a degeneracy between the slope and the scatter of the SMHM when using the SMHM to derive the stellar mass function.  Small halos are more likely to scatter to large stellar masses due to the rapidly rising mass function.  Hence, large scatter requires a steeper SMHM slope in order to reproduce the observed stellar mass function.

Related, it is also possible that low mass halos may be entirely devoid of stars, or contain so few as to be beyond detection \citep[e.g.,][]{OShea2015}. Even if one considers scatter in the SMHM relation, it is common in abundance matching to assume that all halos continue to host galaxies.  In reality, at low halo masses, one halo may contain a faint galaxy, %galaxy with stellar mass of 10$^{5-6}$ M$_{\odot}$, 
while another halo at the same mass is entirely devoid of stars. %Again, the early star formation in dwarf galaxies seems to be tied to the mass accretion history of the halo \citep{BZ2014, Sawala2016}.  
Galaxies that form early are more concentrated \citep{Wechsler2002, Zhao2003}.  For those galaxies with low concentration, reionization may stifle star formation entirely \citep[e.g.,][]{gnedin00, g07, Okamoto2008,Brown2012, Brooks2013, Brown2014, Weisz2014b}.  \citet{Sawala2015} used simulations representative of the Local Group to show that the fraction of halos which are unpopulated (not luminous) increases drastically at halo masses less than $10^8$ M$_{\odot}$, suggesting a plummeting galaxy formation efficiency at these mass scales.

A number of simulators have used baryonic simulations to show that they match the derived SMHM at halo masses above 10$^{10}$ M$_{\odot}$ \citep[e.g.,][]{Brook2012, Aumer2013, FIRE,Munshi13}, but only a few have examined the SMHM  of simulated dwarf galaxies below this mass \citep{Munshi13, Shen2014, Onorbe2015, Sawala2015}.  \citet{Sawala2015} studied a larger population of dwarfs than the other works, tracing halos down to $\sim10^8$ M$_{\odot}$.  They showed that an abundance matching that used only {\it populated} halos leads to placing higher stellar mass galaxies into halos than traditional abundance matching, i.e., it alters the slope of the SMHM relation at low masses.
However, almost all of the these low mass halos were satellite galaxies, which may have earlier formation times than field galaxies and be more luminous at a fixed halo mass \citep{BZ2014, Sawala2016}.  %Only $\sim3$ galaxies below a halo mass of $10^9$ M$_{\odot}$ were central, field galaxies.  %two sources: (1) many of these galaxies are satellites, which are not continuously star-forming, and may be quenched and (2) these satellites have been subject to stripping, whether it be tidal or ram pressure stripping. 
We demonstrate in this work that the characteristic ``bend'' in the SMHM at low halo masses found by \citet{Sawala2015} is a feature that arises due to the inclusion of satellite galaxies in the SMHM relation.

In this paper we address the question: for every low mass dark matter halo, how do stars populate it? %In particular, we focus on how stars populate low mass dark matter halos. 
We do this using a simulated zoomed-in region that contains many dwarfs.  We examine both field dwarfs and their satellites, down to lower masses than previously studied. We show that, for many low mass halos, there are potentially no stars that inhabit them. These halos could have masses of $0$ M$_{\odot}$, and are ``dark halos.''  We both quantify the occupation fraction of dwarf galaxy halos in an $\Lambda$CDM framework as a function of decreasing halo mass, and quantify the scatter in the stellar mass at a given halo mass in the faintest dwarfs.  %To do this, we use a zoomed-in region that contains many dwarfs. 
%and allows us to resolve halos down to a mass of $\sim4\times10^5$ M$_{\odot}$, roughly 3 orders of magnitude lower than ever examined previously.  

By quantifying the occupation fraction, slope, and scatter of the SMHM relation, we provide a tool for modelers to populate low mass halos.  This allows us to predict the stellar mass function of dwarf galaxies.  We show that the stellar mass function increases in slope in the stellar mass range of ultra-faint dwarf galaxies (10$^3$ $\lesssim$ M$_{star}$/M$_{\odot}$ $\lesssim$ 10$^5$).  A number of satellites have recently been discovered in this mass range, thanks to the efforts of the ATLAS \citep{Torrealba2016}, {\it Dark Energy Survey} \citep[DES,][]{DES1,DES2, Kim2015a, Kim2015b, Koposov2015, Luque2017}, {\it Hyper Suprime-Cam Subaru Strategic Program} \citep[HSC-SSP,][]{Homma2017}, {\it Magellanic Satellites Survey} \citep[MagLiteS,][]{Drlica2016}, {\it Survey of the Magellanic Stellar History} \citep[SMASH,][]{Martin2015}, and {\it PanSTARRs} \citep{Laevens2015} surveys.  More ultra-faint dwarfs are likely to be found in the coming years from these same surveys, and when the {\it Large Synoptic Survey Telescope} (LSST) comes online \citep{tollerud08}.  Additionally, searches for satellites around dwarf galaxies may also find new ultra-faint dwarfs \citep{Dooley2016}.  Hence, by studying the low mass end of the SMHM relation, we are able to make predictions about this population of new galaxies that are being discovered.

It is the recent success of simulations in matching dwarf galaxy properties \citep[e.g.,][]{G10, BZ2014, Shen2014} that allow us to undertake this work. At the dwarf galaxy scale, the different slopes between the observed galaxy stellar mass function and the $\Lambda$CDM predicted halo mass function require that galaxies at halo masses below $10^{11}$ M$_{\odot}$ have gas cooling and star formation efficiencies much lower than those of Milky-Way sized galaxies.  While this trend was historically difficult to produce in in cosmological simulations, recent high resolution cosmological simulations that resolve scales on the order of giant molecular clouds can include more realistic models for star formation and feedback, resulting in simulations that can successfully reproduce the observed trends in star formation efficiency \citep{FIRE,Aumer2013,Brook2012,Munshi13,Stinson2013,Simpson2013, G15, Wheeler2015, Christensen2016, Fitts2016}.  %These simulations find that baryonic feedback from stars can also reshape the dark matter density profiles of these small halos, producing shallow or ``cored'' profiles.  Stellar feedback is also key in suppressing the star formation efficiency and thus lowering stellar mass of these halos \citep[e.g.,][]{FIRE,Munshi13,}.  
The success of models in reproducing reliable and accurate dwarf galaxies lies in the ability to be able to resolve the impact of baryonic processes on the interstellar medium and star formation \citep{Munshi14}.  When this happens, the simulations also simultaneously reproduce additional observed trends in dwarf galaxies, such as cored dark matter density profiles \citep{G12,PG14,diCintio2014a, Maxwell2015, Onorbe2015, Read2016} and bulgeless disks \citep{brook11, BC2016}. %There are a multitude of baryonic processes that may contribute to the formation of realistic dwarf galaxies including supernova feedback-driven gas outflows, stellar winds from massive stars injecting energy into the ISM.  
%While $\Lambda$CDM simulations and stellar mass-to-halo mass relations (SMHM) from abundance matching agree  discrepancies arise in lower mass halos.  In particular, stellar to total mass estimates for observed dwarf galaxies are an order of magnitude higher than those inferred from abundance matching in $\Lambda$CDM \citep{moore98b,klypin99,madau08,rashkov11}.  

In this paper we predict the fraction of small dark matter halos that should host galaxies, and what scatter the stochasticity of star formation and mass loss of satellites after infall add to the relationship between stellar mass and halo mass. 
We do so with simulations that, a priori, require no further tuning to successfully match observed properties, including published mass-metallicity relationships \citep{Brooks2007, Christensen2016}, cold gas fractions \citep{Munshi13,Brooks2017} and dark matter profile shapes \citep{G10, Christensen2014a}. The inclusion of realistic feedback models, combined with increasing detail and resolution, is key to understanding the complex interdependency between different types of feedback and the formation of dwarf galaxies as they look today. 
We discuss these simulations in Section \ref{Sims}. In Section \ref{Results} we show that, at the lowest halo masses, there no longer is a strict stellar mass-to-halo mass relationship, with the scatter in stellar mass increasing by an order of magnitude over a two decade decrease in halo mass. We also quantify %the change in cumulative halo number counts in the dark matter only (DMO) simulation, the baryon simulation and for halos that are occupied, quantifying 
the occupation fraction of dark matter halos as a function of declining halo mass.  We quantify the slope and scatter of the SMHM relation.
%In section 3, we examine where the simulations fall in the context of the SMHM relation and present our results on the occupation fraction of dark matter halos  
We summarize our results in Section \ref{Summary}. % we discuss the ramifications of our results. %on the scatter in the SMHM relation, in terms of observed magnitudes of galaxies and in which halos they reside.

%%%%%%%%%%%%%%%%%%%%%%%%%%%%%%%%%%%%%%%%%%%
\section{Simulations}\label{Sims}
%%%%%%%%%%%%%%%%%%%%%%%%%%%%%%%%%%%%%%%%%%%

%\subsection{Code \& Simulation Details}
The simulations used in this work are run with the N-Body + SPH code {\sc ChaNGa} \citep{Menon2015} in a fully cosmological $\Lambda$CDM context using WMAP Year 3 cosmology: $\Omega_0=0.26$, $\Lambda$=0.74, $h=0.73$, $\sigma_8$=0.77, n=0.96. 
{\sc ChaNGa} adopts all the same physics modules as in our previous code, {\sc Gasoline}, but utilizes the {\sc charm}++ run-time system for dynamic load balancing and computation/communication overlap in order to effectively scale up the number of cores.
As described in \citet{Keller2014}, {\sc ChaNGa} has improved its SPH implementation in order to more realistically model the gas physics at the hot-cold interface.  The developers of {\sc ChaNGa} are part of the {\sc agora} collaboration, which will compare the implementation of hydrodynamics across cosmological codes \citep{Kim2014}.

The galaxy sample utilized in this manuscript is selected from a uniform dark matter-only simulation of 25 Mpc per side.  From this volume, a field--like region was selected, representing a cosmological ``sheet.''  
%\textcolor{red}{AB: do I have info on overdensity from HST proposal?} 
It was then re-simulated at extremely high resolution using the ``zoomed-in'' volume renormalization technique \citep{katz93,pontzen08}. The zoom-in technique allows for very high resolution, while accurately capturing the tidal torques from large scale structure that deliver angular momentum to galaxy halos \citep{barnes87}.  These zoom-in simulations have a hydrodynamical smoothing length as small as 6 pc, a gravitational force softening of 60 pc, and equivalent resolution to a 4096$^3$ grid.  Dark matter particles have a mass of 6650 M$_{\odot}$, while gas particles begin with a mass of 1410 M$_{\odot}$, and star particles are born with 30\% of their parent gas particle mass. This sample represents a large number of some of the highest resolution published simulated cosmological dwarf galaxies.  We give this set of simulated galaxies the nickname ``The 40 Thieves.'' 

This set of simulations includes metal line cooling and the diffusion of metals \citep{shen10} as well as  the non-equilibrium abundances of H and He.  We apply a uniform, time-dependent UV field from \cite{HM2012} in order to model photoionization and heating.  Additionally,  we adopt a simple model for self-shielding following the models of \cite{pontzen10}. Star formation is modeled based on a gas density threshold comparable to the mean density of molecular clouds and is similar to that described in \cite{G10}.  Simply, star formation occurs stochastically when gas particles become cold (T$< 10^4$K) and when gas reaches a density threshold of 100 amu/cc.  At this threshold, star formation follows a Schmidt Law, as described in \cite{G10}. \citet{G15} showed that this star formation model ($g3$ in that paper) leads to a bursty but continuous star formation history (SFH), and generated mock color-magnitude diagrams to confirm in observation-space that a dwarf galaxy at the same resolution as the dwarf galaxies in this paper matches observed resolved stellar populations. Models that were bursty, but without an underlying continuous SFH, are not consistent with the observations.  

We adopt the ``blastwave'' supernova feedback approach \citep{stinson06}, in which mass, thermal energy, and metals
are deposited into nearby gas when massive stars evolve into supernovae.  The thermal energy deposited amongst those nearby gas neighbors is 10$^{51}$ ergs per supernova event.  Subsequently, gas cooling is turned off until the end of the
momentum-conserving phase of the supernova blastwave. %which is set by the local gas density and temperature and by the total amount of energy injected, typically $\sim$10 million years. %Equilibrium energy rates are computed from the photoionization code Cloudy \citep{ferland1998}, following \citet{shen10}. A spatially uniform, time evolving, cosmic UV background turns on at $z=9$ and modifies the ionization and excitation state of the gas, following an updated model of \citet{haardtmadau96}.  
Note that, unlike some other sub-grid feedback schemes \citep[e.g.,][]{Oppenheimer2010,Marinacci2014}, this model keeps gas hydrodynamically coupled at all times.  The coupling of 10$^{51}$ ergs of energy into the interstellar medium, combined with the turning off of cooling in the affected gas particles, is designed to mimic the effect of energy deposited in the local
ISM by {\it all} processes related to young stars, including UV radiation from massive stars \citep[see][]{Agertz2013}.

%efficient deposition of supernova energy into the ISM, and the modeling of recurring supernova by the Sedov solution, should be interpreted as a scheme to model the effect of energy deposited in the local ISM by {\it all} processes related to young stars, including UV radiation from massive stars \citep{hopkins11,Wise2012}.  

\begin{figure*}
%\begin{center}
%\includegraphics[width=0.95\columnwidth]{SMHM.eps}
\includegraphics[width=1.80\columnwidth]{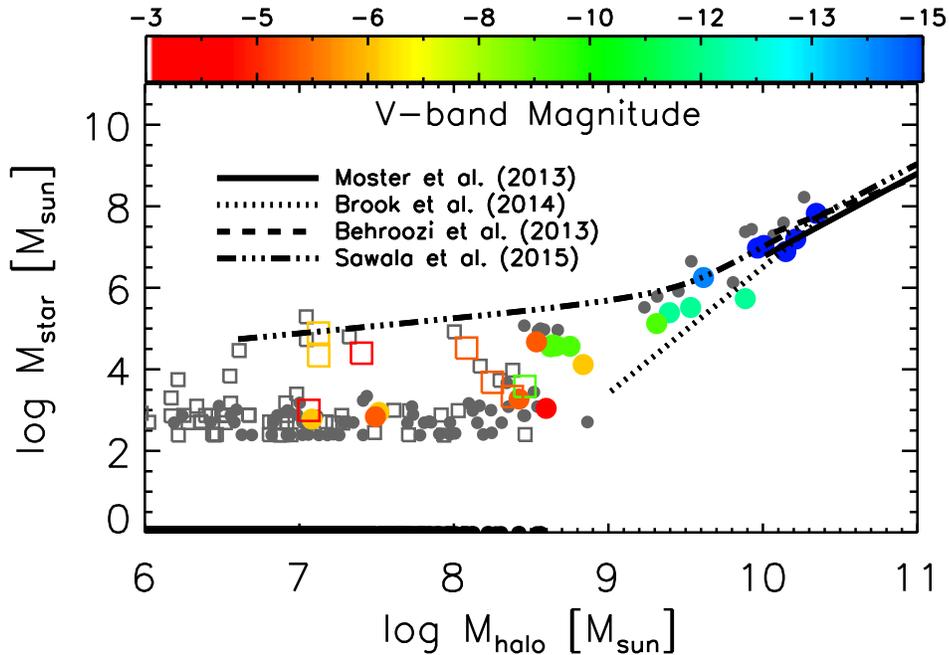}
\caption{{\it Stellar Mass versus Halo Mass.} Results from galaxies in the 40 Thieves simulation are compared to various published SMHM relations. We show well-resolved galaxies (M$_{halo} > 10^7$ M$_{\odot}$) in color, with each galaxy color coded by its absolute $V$-band magnitude, as indicated in the color bar.  For these well-resolved halos, we calculate their stellar mass based on their photometric colors and use the halo mass of their corresponding counterparts in a dark matter-only run \citep[see][]{Munshi13}. The grey points are instead results directly from the baryonic run, and include all halos that host at least one star particle.  The black points along $log$(M$_{star}$) $= 0$ are halos that contain no stars.  Open squares indicate satellite galaxies, while filled circles represent central galaxies. Our simulations are consistent with all relations at halo masses above 10$^{10}$ M$_{\odot}$, but the scatter in the relation increases with decreasing halo mass.  Below $\sim10^9$ M$_{\odot}$, there is no single stellar to halo mass relationship for the simulated dwarfs.    %The grey points are instead direct from the baryonic run results, and include all halos that host at least one star particle.  The black points along $log$(M$_{star}$) $= 0$ are halos that contain no stars.  {\bf Open squares indicate satellite galaxies, while filled circles represent central galaxies.}
}
%\end{center}
\label{fig:smhm}
\end{figure*}

Individual halos are identified using {\sc Amiga's Halo Finder}\footnote{AHF is available for download at http://popia.ft.uam.es/ AHF/Download.html.} \citep[AHF,][]{gill04,knollmann09}. The virial radius is defined to be the radius for which the average halo density is some multiple of the background density, following an overdensity criterion that varies with redshift \citet{bryannorman}. At $z=0$, the virial radius is defined as the radius within which the average halo density is approximately equal to 100 times the critical density. For satellites in this work, we trace back the main progenitor to find the maximum halo mass that the halo attained.  At each snapshot, the main progenitor is defined to be the halo in the previous step that contains the majority of the particles in the current halo.
 
%%%%%%%%%%%%%%%%%%%%%%%%%%%%%%%%%%%%%%%%%%% 
\section{Results}\label{Results}
%%%%%%%%%%%%%%%%%%%%%%%%%%%%%%%%%%%%%%%%%%%

\begin{figure*}
%\begin{center}
%\includegraphics[width=0.95\columnwidth]{SMHM.eps}
\includegraphics[width=1.70\columnwidth]{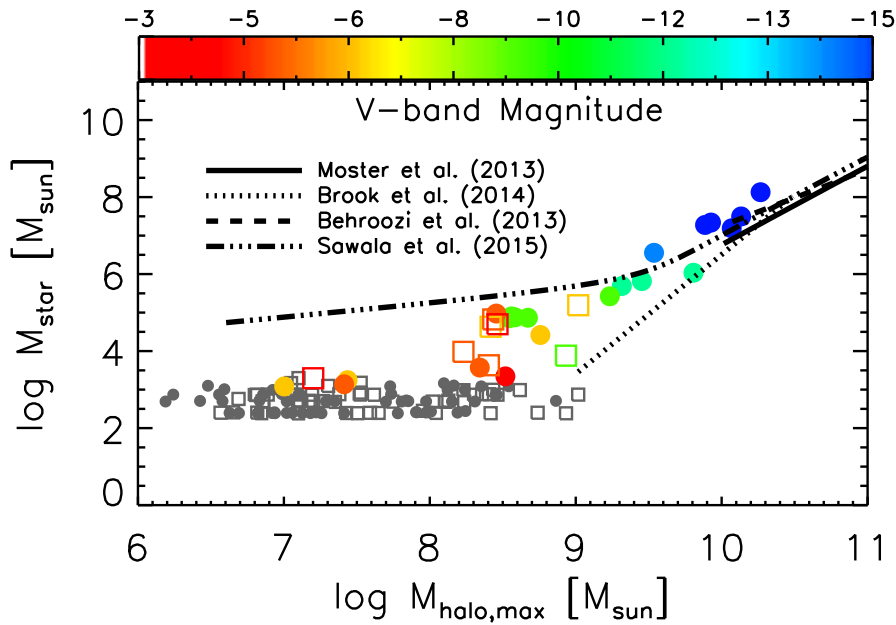}
\caption{{\it Stellar Mass versus Maximum Halo Mass.} Similar to Figure \ref{fig:smhm} except that satellite galaxies at $z=0$ are now shown at the time of their maximum halo mass, and the stellar mass at that corresponding time.  Central galaxies are as in Figure \ref{fig:smhm}, except that all results are now taken from the baryonic simulations (e.g., we do not use the corresponding dark matter-only halo mass nor do we use photometrically derived stellar masses). Again, grey points include all halos, even those that are not well resolved, and open squares indicate satellite galaxies, while filled circles represent central galaxies. Compared to Figure \ref{fig:smhm}, it is clear that adopting the maximum halo mass decreases the scatter at the low mass end.
}

%\end{center}
\label{fig:mmax}
\end{figure*}

In Figure \ref{fig:smhm}, we show the stellar mass of the simulated galaxies in the 40 Thieves as a function of halo mass.  Also plotted are various published SMHM relations that were derived at higher masses \citep{Moster2013,Brook2014,Behroozi2013,Sawala2015}.  In this Figure, the colored points are only those halos with M$_{halo} > $10$^7$ M$_\odot$ and M$_{star} > $600 M$_\odot$.  This halo mass corresponds to the hydrogen cooling limit, and focusing on halos greater than this mass ensures that our dark matter halos are sufficiently well-resolved (more than 1500 particles).  For the colored points, the stellar masses in Figure \ref{fig:smhm} are calculated using each simulated galaxy's photometric color, as described in \citet{Munshi13}.  The plotted halo masses are taken from a dark matter-only simulation of the same galaxies in order to be consistent with the fact that the derived SMHM relations that we compare to at the high mass end used mass functions from dark matter-only simulations.  The small grey points are instead the results for the same halos if the values are taken from the baryonic runs, summing both the stellar and dark matter particle masses, and also including all halos with at least one star particle, regardless of halo mass.  Open squares indicate satellite galaxies, while filled circles represent central galaxies. The black points along $log$(M$_{star}$) $= 0$ are halos that contain no stars.

Simulated galaxies shown as colored points in Figure \ref{fig:smhm} are color coded by their $V$-band magnitude.\footnote{The $V$-band magnitude is calculated in {\it pynbody} \citep{pynbody}, which utilizes the Padova simple stellar population models \citep{Marigo2008, Girardi2010} found at http://stev.oapd.inaf.it/cgi-bin/cmd.}  This color coding helps to emphasize that the 40 Thieves sample includes a wide range of stellar populations and star formation histories, from bright dwarfs to ultra-faint dwarfs.  

At the faintest end of the SMHM relation probed by the 40 Thieves, a galaxy cannot be assigned a unique halo mass based solely on its luminosity.  %Abundance matching, in practice, typically assigns a single halo mass at a fixed luminosity.  While this may work at higher masses, Figure \ref{fig:smhm} highlights that this assumption is no longer valid at the lowest masses.  For ultra-faint dwarf galaxies, assigning a dwarf a particular halo mass based on its stellar mass, or its $V$-band magnitude, is incredibly difficult, e.g., galaxies with $M_V \sim -4$ reside in halos that span two orders of magnitude in mass.  
For example, as shown in Figure \ref{fig:smhm}, a galaxy as bright as $M_V \sim -7$ can be hosted by halo masses that span roughly two orders of magnitude, ranging from $\sim$10$^7$  M$_\odot$ to $\sim$10$^9$ M$_\odot$.  This is in stark contrast to conventional abundance matching, which assumes there exists a one-to-one relationship between luminosity (or stellar mass) and halo mass.

Figure \ref{fig:smhm} demonstrates that, as halo masses decline, the scatter in the stellar masses of the 40 Thieves drastically increases. %As we discuss below, this scatter is a lower limit due to our restriction of using M$_{halo} >$10$^7$ M$_\odot$. 
However, Figure \ref{fig:mmax} emphasizes that the scatter at the lowest masses (M$_{halo} \sim 10^7$ M$_\odot$) can be attributed to including satellites of the dwarf galaxies in the relation.  Figure \ref{fig:mmax} shows the maximum halo mass of satellites (typically at a time just before infall onto their parent halo), and the stellar mass at that time.   Similarly, the $z=0$ SMHM relation of \citet{Sawala2015} is dominated by satellite galaxies at the low mass end.  

Although the most massive central dwarf in our simulation is comparable in mass to the SMC, it has one satellite with M$_{star} \sim$1000 M$_\odot$.  In fact, the six most massive central dwarfs in the simulation each contain at least one satellite brighter than M$_V = -4$, while one galaxy has two, for a total of eight well-resolved satellites shown in Figures \ref{fig:smhm} and \ref{fig:mmax}. The stellar masses (halo masses) of the parent dwarf galaxies range from 10$^6$ -- 10$^8$ M$_\odot$ (6$\times$10$^9$ -- 2$\times$10$^{10}$ M$_\odot$). %The number of satellites increases to 55 if we include all satellites with at least 1 star particle.   }

Considering the maximum halo mass for the satellites reduces the scatter in stellar mass at the lowest halo masses, from $\sim$1 dex to $\sim$0.25 dex at M$_{halo} \sim 10^7$ M$_{\odot}$. In some cases subhalos lose more than an order of magnitude in halo mass after infall. The halos with M$_{star} \sim 10^4$ M$_\odot$ and M$_{halo} \sim 10^7$ M$_{\odot}$ in Figure \ref{fig:smhm} are satellites that originally had M$_{halo} > 10^8$ M$_{\odot}$ and have been substantially stripped of mass after infall.  Stellar masses are more robust, and only those halos with significant halo stripping have lost about a factor of two in stellar mass.  This is consistent with earlier findings that $\sim$90\% of the dark matter mass can be stripped before stars are stripped \citep{Penarrubia2008, Libeskind2011, Munoz2008, Chang2013, BZ2014}. %[see][and references therein]

\subsection{Quantifying the Scatter}

In this section we quantify the scatter in our SMHM relation.  We use the $z=0$ results for halo masses of the satellites.  Previous constraints on the low mass end of the SMHM relation use galaxies drawn from the Local Volume to derive a stellar mass function \citep{Brook2014, GK2014}, including satellite galaxies.  Hence, the adopted halo mass functions should not discriminate between satellites and centrals. 

\begin{figure*}
\includegraphics[width=1.0\columnwidth]{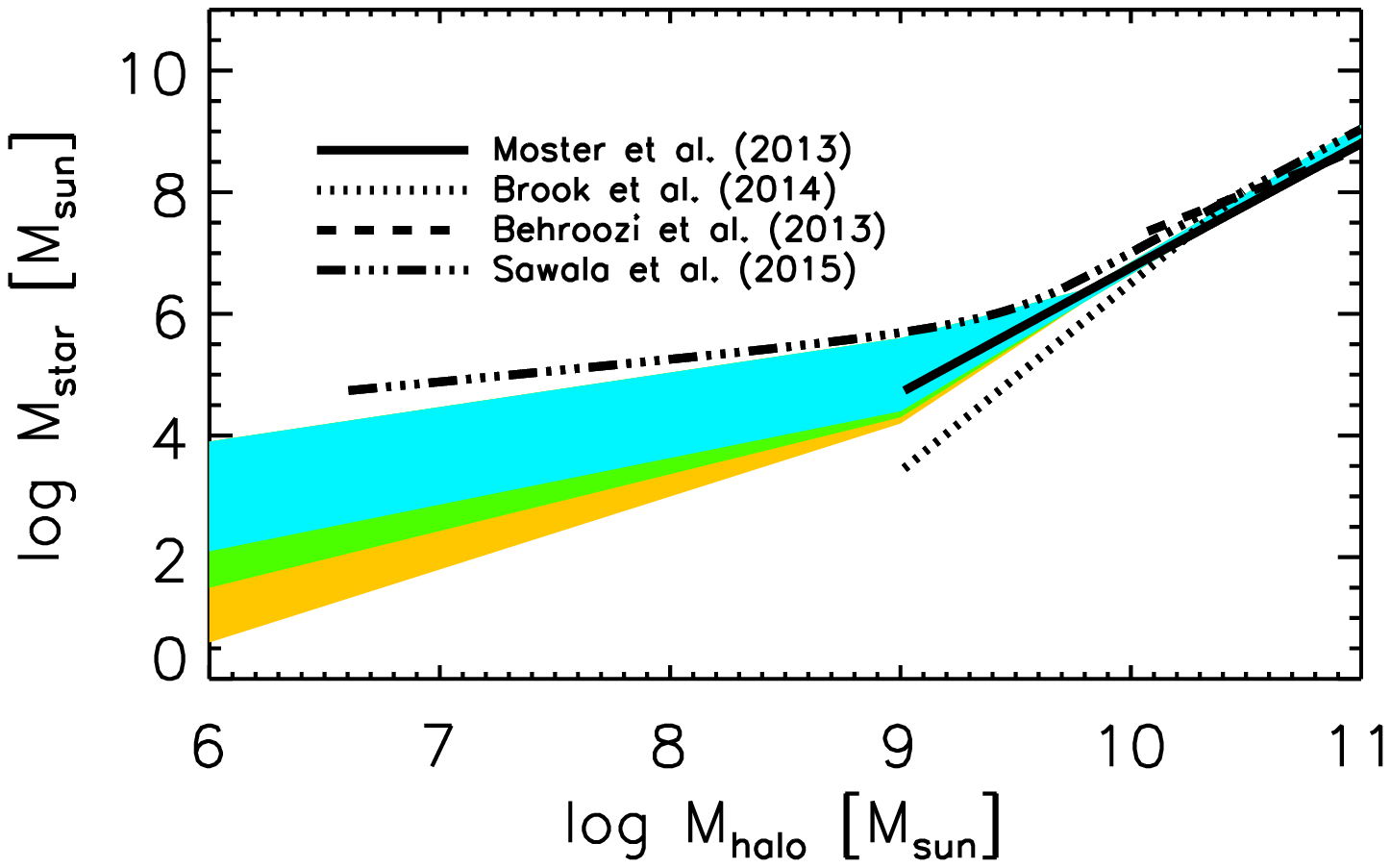}\includegraphics[width=1.0\columnwidth]{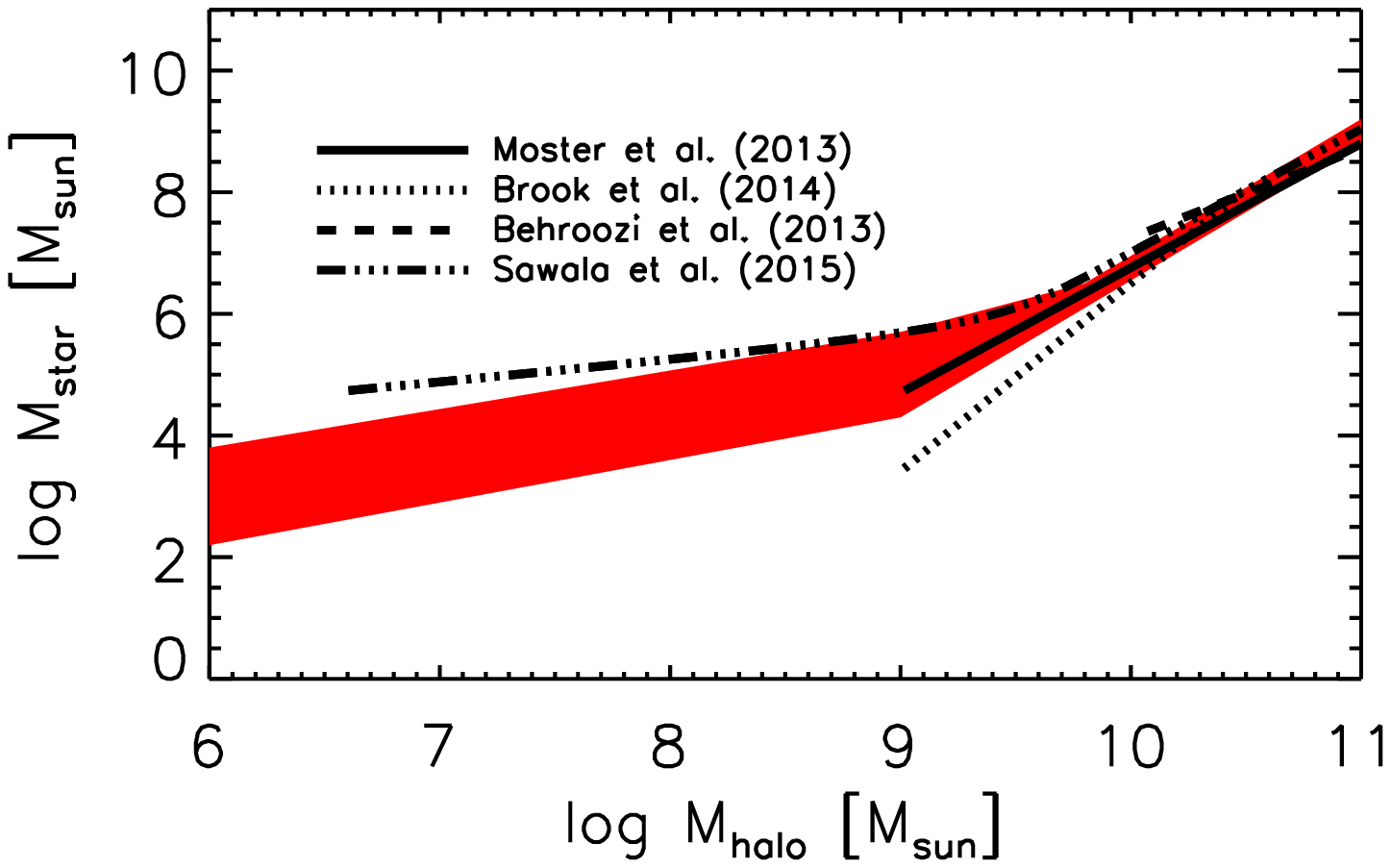}
\caption{{ \it Left Panel: Scatter in stellar mass at a given halo mass at $z=0$ with various stellar mass cuts.  Right Panel: Scatter in halo mass at a given stellar mass.} Lines are a sample of published SMHM relationships.  Colored shaded regions are bounded by mean scatter in a given mass bin. Cyan shading represents scatter from well-resolved halos, green shading represents the increased scatter from halos that contain one star particle or more, and orange shading represents increased scatter by including dark halos. %
}
\label{fig:scatter}
\end{figure*}

\subsubsection{Well-resolved Halos}

Figure \ref{fig:scatter} shows the range of scatter in our simulated SMHM relation.  We adopt log-normal scatter.  The left plot shows the scatter in $log(M_{star})$ at a fixed halo mass, while the right plot shows the scatter in $log(M_{halo})$ at a fixed stellar mass. 
The cyan band in the left panel shows the scatter for all of the well-resolved halos that are shown in Figure \ref{fig:smhm} (below we discuss the scatter when all luminous halos are included, and when dark halos are included as well).  
At a given halo mass, the scatter in stellar mass is of order 0.1 dex for halos more massive than $\sim10^9$ M$_{\odot}$.  The scatter in stellar mass increases to 0.9 dex at the smallest halo masses for these well-resolved halos.  %With this scatter, there is no longer a single stellar to halo mass relationship for the faintest dwarf galaxies.  
In the right hand panel, the scatter in halo mass at a given stellar mass is shown by the red band. For stellar masses greater than 10$^{5.5}$ M$_\odot$, the scatter in halo mass is of order 0.2 dex, while the scatter is 0.7 dex at lower stellar masses.  %On this plot, we show all halos up until our mass resolution limit: the smallest halos here contain at least 1500 DM particles and at minimum 8 star particles.

%\textcolor{red}{Discuss Shea's article.  Note $\alpha$ = 2.4 (consistent with Moster at high mass end), slope $\gamma$ = -0.26}
As noted in the Introduction, \citet{GK2016} recently demonstrated the impact of scatter in the SMHM relation on the resulting stellar mass function.  They explored both a model in which the scatter is constant as a function of halo mass, and a model in which the scatter increases as halo mass declines.  Clearly, our results favor a model in which the scatter increases toward low halo masses.  \citet{GK2016} quantify the increasing scatter as follows:
\begin{equation}
\sigma = 0.2 + \gamma (\rm log_{10} M_{halo} - \rm log_{10} M_1)
\label{eq1}
\end{equation}
where $\sigma$ is the scatter, $\gamma$ is the rate as which the scatter grows, and M$_1$ is a characteristic mass above which the scatter remains constant. \citet{Munshi13} demonstrated that our simulations are in good agreement with the derived SMHM of \citet{Moster2013} at the high mass end, i.e., a slope of the SMHM $\alpha = 2.4$.  This slope continues to fit our results down to $log$(M$_{halo}$) $= 8.4$.  Below this halo mass, we must define a new, shallower slope in order to fit the trend of the well-resolved halos, $\alpha = 0.64$.  Our scatter remains roughly constant (and small) above M$_1 = 5\times10^9$ M$_{\odot}$, and grows below this halo mass.  Using Eqn. \ref{eq1} with the well-resolved simulated galaxies yields $\gamma = -0.26$.  

\subsubsection{All Populated Halos}

We next loosen our restriction from all well-resolved halos to halos that contain one or more star particles.  The grey points in Figure \ref{fig:smhm} show results for all halos in the simulation that contain stars.  Note that in this case, we drop our calculation of stellar masses according to photometric color, and our assignment of halo mass based on the dark matter-only run.  That is, the grey points in Figure \ref{fig:smhm} are direct from our baryonic simulation results, including for the halos that were also in our well-resolved sample. Our stellar resolution limit defines the lower bound (our lowest mass star particles are 216  M$_{\odot}$).  With the caveat that we are now likely incomplete, we repeat the fit including all halos with stars and find that the slope of the scatter from Eqn. \ref{eq1} increases to $\gamma = -0.48$.  This increased scatter is shown by the green band in the left panel of Figure \ref{fig:scatter}.

\begin{figure*}
\includegraphics[width=0.95\columnwidth]{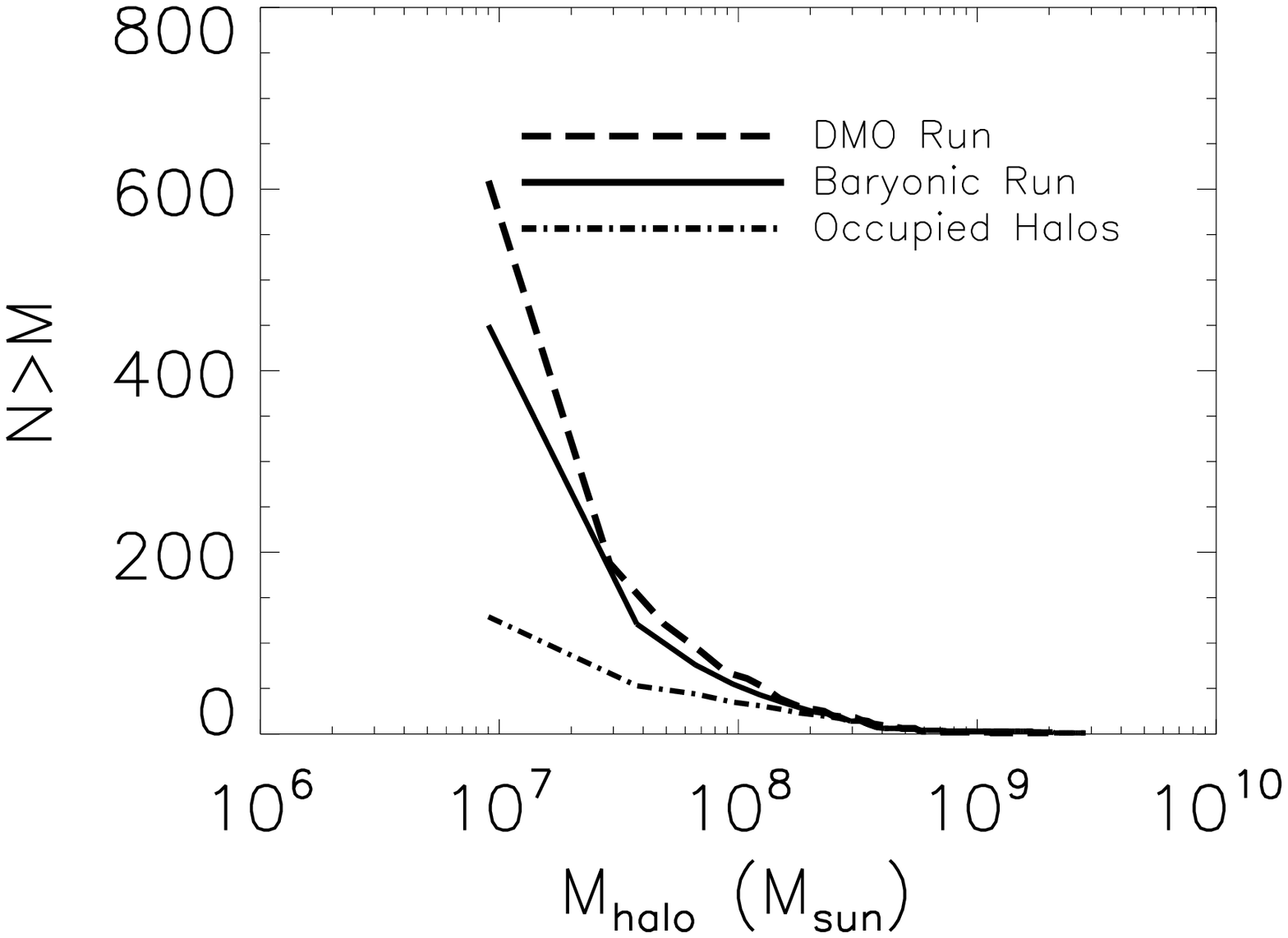}\includegraphics[width=0.95\columnwidth]{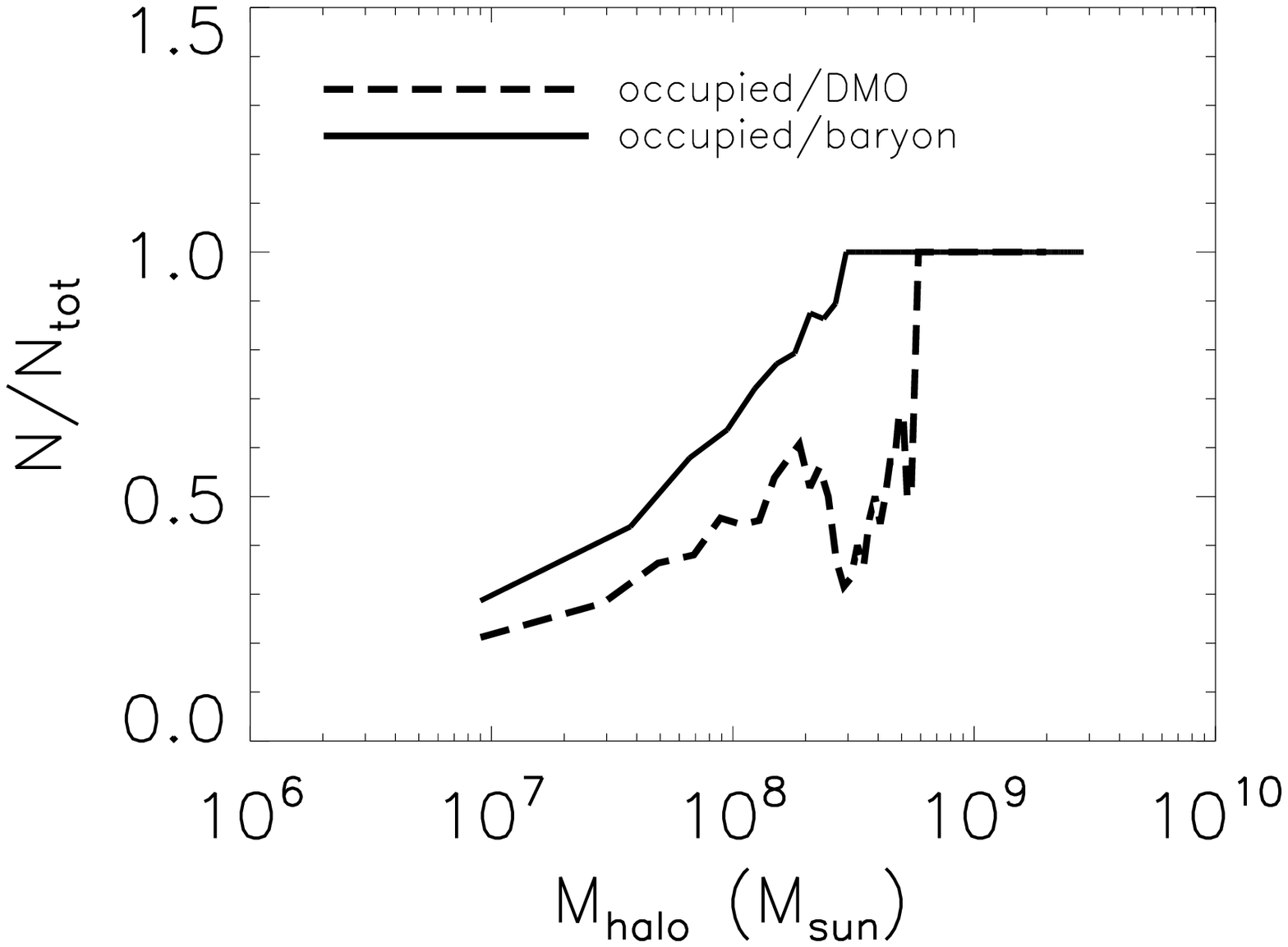}
\caption{{\it Cumulative halo mass functions.} Left Panel: The dashed line is the cumulative mass function for the dark matter-only (DMO) run; the solid line is the cumulative mass function for the baryonic run using all halos; the dot-dashed line the cumulative mass function using only occupied (luminous, with a minimum of one star particle) halos in the baryonic run. Right Panel: The ratio of the cumulative number of occupied (luminous) halos relative to the corresponding number of halos in the DMO run (dashed line) and the baryon run (solid line).  The fraction of dark (non-luminous) halos increases with decreasing halo mass.
  %The number of ratio declines rapidly for decreasing halo mass, with the majority of halos remaining "dark" at $10^7 M_{\odot}$%
}
\label{fig:dark}
\end{figure*}

\subsubsection{All Halos}

Next we extend the discussion to include dark halos.  In Figure \ref{fig:smhm} we show dark halos by arbitrarily setting their stellar mass to 1 M$_{\odot}$, so that they pile up along $log$(M$_{star}$) $= 0$.  We quantify the fraction of halos that are occupied by any stars (i.e., contain one or more star particles) as a function of halo mass in Figure \ref{fig:dark}.  We note that baryonic physics can impact halo mass, lowering the final M$_{halo}$ of a given galaxy in the baryonic run from the corresponding halo run with dark matter only.  As was mentioned above, the well-resolved halos in Figure \ref{fig:smhm} use halo masses from a dark matter-only run for consistency with derived SMHM relations; Figure \ref{fig:dark} demonstrates why.  The left plot of Figure \ref{fig:dark} shows the cumulative halo mass functions for both the baryonic and dark matter-only versions of the 40 Thieves.  The solid and dashed lines trace all halos, regardless of whether they contain stars.  It can be seen that the mass functions of the two runs do not quite match.  This mismatch is due to the fact that in the baryonic run 
the ejection of baryons from low mass halos (either by heating from the UV background or as a result of supernova feedback) reduces the growth rate of a halo.  The overall effect is that a given halo in the baryonic run is less massive than in the corresponding dark matter-only simulation \citep[see also][]{Sawala2013,Munshi13}.  %In \citet{Munshi13} we address the difference in halo mass between DMO and baryon simulations which supports results from \citet{Sawala12,Sawala15}.  
As a direct result, the total number of halos with M$_{halo} > $10$^7$ M$_\odot$ is reduced ($\sim$75\%) in the baryonic run compared to the dark matter-only run. %in our lowest mass bin. 
Figure \ref{fig:dark} also shows the cumulative mass function for only those halos that ``host a galaxy,'' i.e., contain a minimum of one star particle \citep[see also][]{Sawala2015}. % For this mass function, we include all halos that have a minimum of one star particle \citep[see also][]{Sawala2015}.

%In Figure 2, we specifically look at the abundance of halos as a function of halo mass between both iterations of the simulation: the baryonic simulation and the DMO simulation- the difference between these two cumulative halo functions is a direct result of the ejection of baryons lowering the total halo mass halos.  This effect was also demonstrated in \citet{Munshi13}. In line with previous works, Figure 2 also shows that baryonic physics affects the cumulative number of halos as a function of halo mass in the low mass regime: there are fewer ($\sim$85\%) of faint halos in the baryon run compared to the DMO run in our lowest mass bin.  

%Our simulation reproduces the trend presented in \citet{Sawala15}, but we do so while {\it simultaneously} matching published SMHM relationships \citep{Moster12,Behroozi13,Brook14}, as in \citet{Munshi13} and shown here in Figure 1.  We do not assume all halos to be populated, and calculate our stellar and halo masses as in \citet{Munshi13} to produce Figure 1.
%The total number of halos with M$_{halo} >$ 10$^7$ M$_\odot$  is $\sim$25\% that of the dark matter-only run.

In the right panel of Figure \ref{fig:dark}, we plot the ratio between the number of occupied halos (that contain one or more star particle) to the total number of halos in both the dark matter-only and baryonic simulations. At halo masses of $10^7$ M$_\odot$, only 20\% of halos are occupied compared to the number of halos in the dark matter-only run.  Due to the fact that the baryonic run contains fewer halos at $10^7$ M$_\odot$, the fraction is slightly higher relative to the baryonic run; luminous halos compose 30\% of the total number of halos with masses above $10^7$ M$_\odot$ in the baryonic run.  %The dark halos in the baryon run form no stars up until our mass resolution limit and 
%We verified that our dark halos are not just dark at $z = 0$; their baryon fractions never exceed 2\% throughout their whole history.

Hence, in the presence of baryons, $\Lambda$CDM predicts a significant number of dark halos at very low mass. We attempt to recalculate the scatter and slope of the SMHM when these halos with M$_{star} = 0$ M$_\odot$ are considered. We do this not to redefine the definition of a galaxy \citep{Willman2012}, but purely as a demonstration of how the scatter in the relationship grows if dark halos are included. We note that log-normal scatter is technically no longer a good description of the relation, as there are more dark halos than luminous halos below M$_{halo} \sim 10^8$ M$_\odot$, and the scatter is no longer evenly distributed above and below the mean.  However, if we ignore that fact and use Eqn. \ref{eq1} including all halos, we find that the rate at which scatter grows increases drastically, resulting in a slope $\gamma = -0.85$.  The slope at the low mass end of the relation also steepens below $log$(M$_{halo}$) $= 8.4$, to $\alpha = 0.84$. 
The increased scatter including the dark halos is shown in the left hand panel of Figure \ref{fig:scatter} by the yellow band.  As with the well-resolved halos, the scatter above halo masses of 10$^9$ M$_\odot$ remains 0.1 dex, as there are no dark halos in this mass range.  The scatter increases to 0.8 dex at a halo mass of 10$^8$ M$_\odot$, and to 2.4 dex at a halo mass of 10$^7$ M$_\odot$.

\subsection{Populating a Halo Mass Function}

\begin{figure*}
%\begin{center}
\includegraphics[width=0.95\columnwidth]{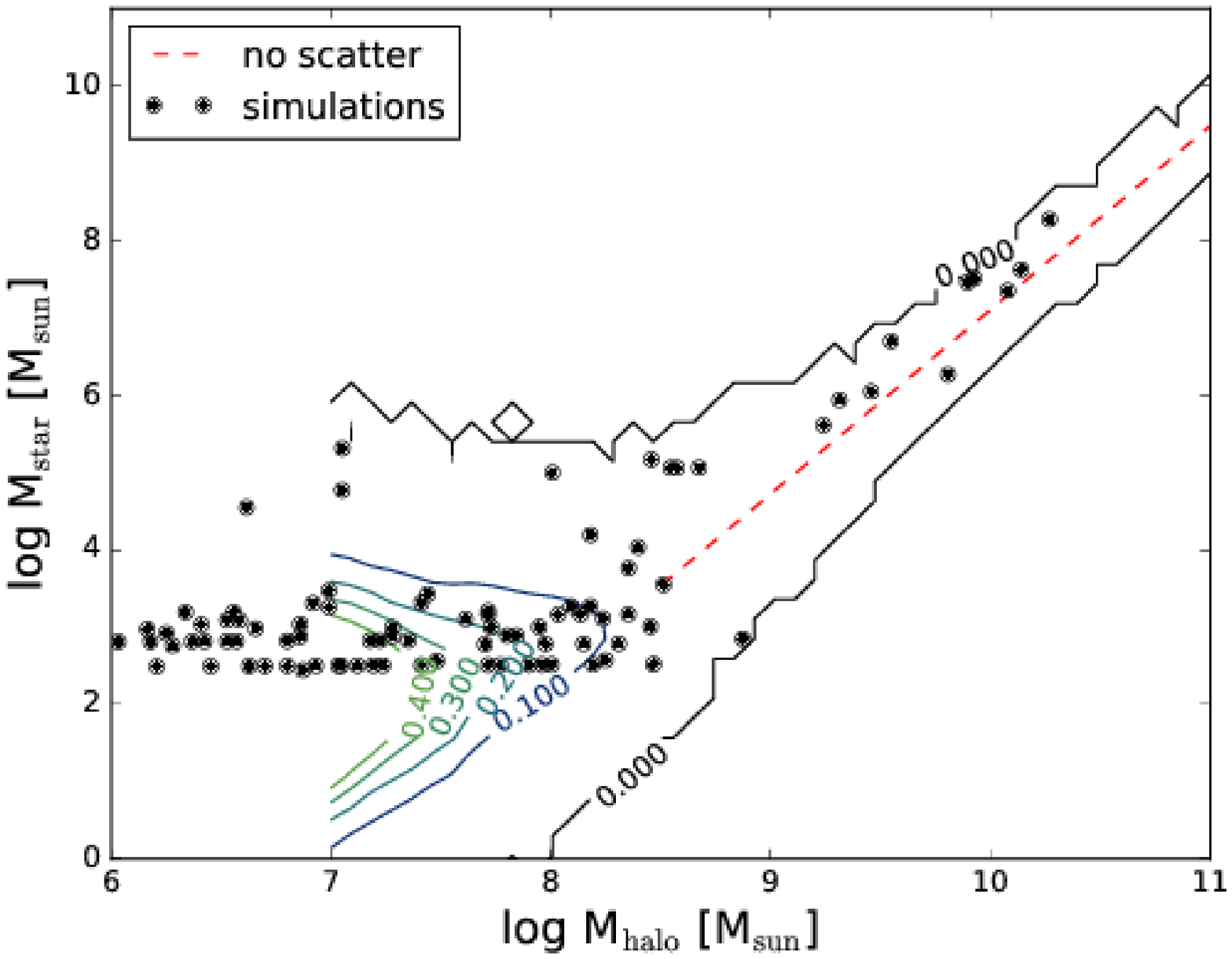}\includegraphics[width=0.95\columnwidth]{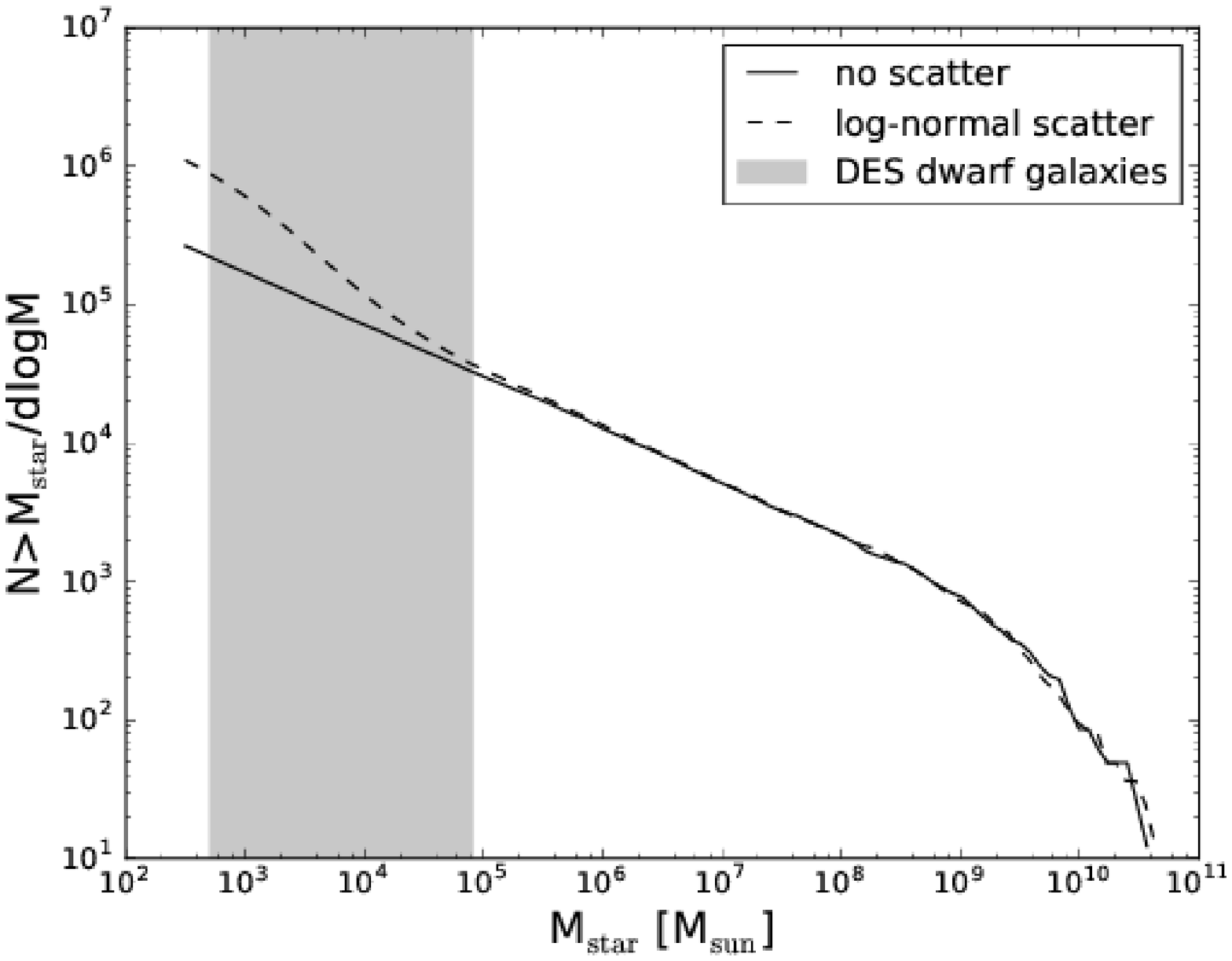}
\caption{{\it Populating a Halo Mass Function Including Scatter.} Left Panel: The black points are the same as the grey in Figure \ref{fig:smhm}.  The constant-density contours are the 2D probability density function resulting from stochastically populating 200,000 halos with log-normal scatter below $log$(M$_{halo}$) $= 11.5$, while requiring that the fraction of dark halos (i.e., those below our resolution limit of 1 star particle) reproduces the results in the right panel of Figure \ref{fig:dark} as a function of halo mass. We adopt the trends defined for our well-resolved halos: a slope of the SMHM $\alpha = 2.4$ for 8.4 $< log$(M$_{halo}$) $< 11.5$ (shown by the red dashed line), with a shallower slope of $\alpha = 0.64$ at lower halo masses.  The scatter is fixed to a constant 0.2 above M$_1 = 5\times10^9$ M$_{\odot}$, and linearly grows below this halo mass so that $\gamma = -0.26$. Adopting these criteria generally reproduces our SMHM relation results.  
Right panel: The change in the stellar mass function including scatter (dashed line).  We compare to the case with no scatter (solid line).  The scatter is small enough at high halo masses to have almost no impact on the resulting mass function.  It is not until the scatter grows to nearly 1 dex at the smallest halo masses (10$^7$ M$_{\odot}$) that the scatter begins to alter the stellar mass function.  This change falls in the range of masses of the recently discovered DES dwarfs (shaded region).}
%\end{center}
\label{fig:elaad}
\end{figure*}

We demonstrate in this subsection how to reproduce our results given a halo mass function.  The contours in the left panel of Figure \ref{fig:elaad} shows the results of stochastically populating 200,000 halos below $log$(M$_{halo}$) $= 11.5$ given a halo mass function in our WMAP3 cosmology.  We wish to reproduce the general trends seen in the black points, which are the same as the grey points in Figure \ref{fig:smhm}, but we also assume there may be real galaxies that host stellar populations below our resolution limit of one star particle.  To populate our halos, we adopt the log-normal scatter for the well-resolved halos, but we simultaneously require that the fraction of luminous galaxies (i.e., the fraction of galaxies with stellar masses above our resolution limit) matches the results in the right hand panel of Figure \ref{fig:dark}.  Recall that for our well-resolved halos, a slope of the SMHM $\alpha = 2.4$ is the best fit for 8.4 $< log$(M$_{halo}$) $< 11.5$, with a shallower slope of $\alpha = 0.64$ at lower halo masses.  The scatter is constant at 0.2 dex above M$_1 = 5\times10^9$ M$_{\odot}$, and linearly grows below this halo mass, i.e., $\gamma = -0.26$.  Once populated in halo and stellar mass, the points are binned into a 2D histogram, which is normalized such that the total volume of the histogram is unity. Resulting contours of constant density from this histogram are plotted in Figure \ref{fig:elaad}.

The right panel of Figure \ref{fig:elaad} shows the resulting stellar mass function after populating 200,000 halos.  \citet{GK2016} recently demonstrated that large scatter impacts the observed stellar mass function.  Due to the rapidly rising halo mass function, halos are more likely to scatter to larger stellar masses, an effect that is increasingly noticeable as scatter increases.  Our scatter grows in such a way that it only reaches relatively large scatter ($\sim$1 dex) at very low halo masses that host primarily ultra-faint galaxies.  Thus, the scatter only has a noticeable impact on the stellar mass function at masses below $\sim$10$^4$ M$_{\odot}$.  These are luminosities below which the observed luminosity or stellar mass function are not currently complete, but are being or will be probed by surveys like DES, HSC-SSP, or with LSST \citep{tollerud08, Walsh2009}.  In the right panel of Figure \ref{fig:elaad}, we shade the mass region of the new dwarf satellites that have been found in the first two years of the DES \citep{DES1, DES2}.

%{\bf Additionally, some work has found that small galaxies are more vulnerable to stripping \citep[e.g.,][]{Fillingham2016}. However, if low stellar mass galaxies live in halos of mixed mass, the implications for quenching timescales become more complicated. In particular, if stripping drives quenching down to the lowest masses, like those probed in this paper, this has broad implications for the CGM around dwarf galaxies: namely the implication that the CGM around dwarfs is clumpy and dense ($\sim${2-20} times the mean density).}

Despite the fact that the stellar mass function is not altered in the current range that is observationally complete, the stochastic nature of star formation in low mass dwarfs is still significant, leading to an increasingly steep slope at soon-to-be probed masses.  \citet{tollerud08} estimates that such faint dwarf galaxies should be observed out to $\sim$1 Mpc by LSST after the full co-added data are collected.  Thus, our predicted increase in the slope of the stellar mass function can be tested.

\subsection{Star Formation Model}

The star formation prescription used in this work adopts a high density threshold for star formation ( $\rho > $ 100 amu/cc, T$< 10^4$K), which has been shown to produce resolved stellar populations as seen in real dwarfs \citep{G15}.  However, \citet{Christensen2014a} demonstrated that gas that is restricted to forming from molecular (H$_2$) gas forms at even higher densities and colder temperatures.  This change does not affect our SMHM results for halos more massive than $\sim$10$^8$ M$_{\odot}$, because our results in this work at the high mass end are in agreement with \citet{Munshi13}, who examined the SMHM using simulated galaxies in which stars formed only in the presence of H$_2$ \citep{Christensen2012}.  

However, it is not clear that our results at lower masses would also remain unchanged if we adopted an H$_2$-based star formation prescription, particularly for the halos that form only one star particle.  Current observations are incomplete in the ultra-faint dwarf galaxy luminosity range. %This is also the halo mass range that is observationally incomplete, containing ultra-faint dwarf galaxies.  Thus, there are no current observational constraints in this mass range.  
If a change in the star formation prescription yields different results at the low mass end, this could result in a different prediction for the stellar mass function.  This is unlikely, as our predictions are based off of the scatter in our well-resolved halos, which also form stars at the high densities achieved when H$_2$ star formation is followed.  

Despite these caveats, in a future work we will examine the differences in the SMHM relation at lower masses that result from restricting stars to forming in even higher density gas (Munshi et al., {\it in prep}).  Studying differences in star formation in the lowest mass halos has multiple implications.  First, any difference in the stellar mass function between the two models means that the observed stellar mass function from LSST at ultra-faint dwarf scales can constrain star formation models.  Second, changes in star formation at the lowest masses will alter predictions for how many faint satellites will be observed around dwarf galaxies \citep{Sales2013, Wheeler2015, Dooley2016, Dooley2017}.  %Not only will we make predictions for the stellar mass function in the two models, as a test of the star formation model that can be constrained by future surveys, we also will discuss the liklihood of finding observable dwarf galaxy satellites as a function of star formation model in the local group.
We will make predictions for each of these observations in the two different star formation models (Munshi et al., {\it in prep}).

%{\bf Maybe discuss here that it's not clear if molecular gas is required, or if molecular gas is just a tracer of dense gas?}  

\section{Summary}\label{Summary}
In this paper, we use a large sample of extremely high resolution simulated dwarf field galaxies (SMC-mass and smaller) and their satellites in order to predict the SMHM relation at low halo masses. %This simulation includes a new SPH implementation that addresses the traditional problems with SPH, a blastwave subgrid scheme for supernova feedback, star formation based on a high gas-density threshold, metal dependent radiative cooling and the turbulent diffusion of metals and energy.  In particular, 
This is the first prediction of the SMHM relation for field galaxies down to M$_{halo} = 10^7$ M$_\odot$, and the first time that the scatter at the low mass end has been robustly quantified.  %We quantify the scatter in the relation, first by considering only our simulated halos that are luminous, and next by considering the true scatter when dark halos (that contain no stars) are considered.  
We demonstrate that (1) derived SMHM relations cannot be extrapolated from higher masses to lower masses, and thus (2) the halo mass of a faint dwarf galaxy cannot be inferred {\it a priori}. 

We quantify the scatter both in terms of scatter in halo mass given a stellar mass and scatter in stellar mass for a given halo mass.  We show that scatter increases dramatically for lower-mass halos.  When we consider only well-resolved simulated halos above $10^7$ M$_\odot$ (i.e., those that contain more than 1500 particles and that are above the hydrogen cooling limit), the scatter in M$_{star}$ at M$_{halo}$ $\sim10^7$ M$_{\odot}$($z=0$) reaches values as large as $\sim1$ dex.  Assuming log-normal scatter, the rate of increase in scatter is quantified by $\gamma$, where $\gamma = -0.26$ for well-resolved halos.

 For the first time, we also demonstrate that much of this scatter is due to the inclusion of satellite galaxies in the SMHM relation.  For our well-resolved halos, considering the maximum halo mass achieved by our satellites decreases the scatter in the SMHM from $\sim1$ dex to $\sim0.25$ dex at a halo mass of 10$^7$  M$_{\odot}$.  The origin of this scatter is the fact that the satellites of our dwarf galaxies can lose substantial mass after infall, including satellites that lose an order of magnitude or more in halo mass.

The substantial stripping that some of our satellites of dwarfs experience may run counter to naive expectations that tidal mass loss is inconsequential in the shallow potential wells of dwarf galaxies.  In a future work (Ahmed et al., {\it in prep}), we will examine the mass loss of these dwarf galaxy satellites in more detail.  It has been suggested that some of the recently discovered DES dwarfs may be satellites of the Magellanic Clouds \cite{Donghia2008, Deason2015, Yozin2015, Jethwa2016, Dooley2017, Sales2017}, as associations of dwarf galaxies and their satellites are a natural outcome of CDM \citep{Stierwalt2017}, and could be responsible for many of the ultrafaint dwarfs in the MW halo.  None of these works have highlighted that the satellites of dwarfs may be substantially stripped, which has implications for their expected velocity dispersions. 

However, the majority of halos with $10^7 <$ M$_{halo}$/M$_\odot < 10^8$ have no galaxy associated with them.  They are dark halos.  %halos that are either too poorly populated by stars and gas to be observed by present-day standards, or that contain any stars or dark matter. 
Using the strictest definition possible (no star particles formed in the simulated halo), we derive the fraction of dark halos at a given halo mass.  %We then recalculate the scatter of the SMHM by including these halos with M$_{star} = 0$ M$_\odot$.  The scatter in stellar mass at M$_{halo} = 10^7$ M$_{\odot}$($z=0$) increases to 2.4 dex.  In this case, $\gamma = -0.85$. 
By including our estimate of the fraction of dark halos and the scatter of our well-resolved halos, we have provided a tool that can be used to stochastically populate theoretical models at low masses instead of relying on abundance matching.  We demonstrate this by populating a halo mass function including the scatter for our well-resolved halos, and requiring that we reproduce the fraction of dark halos in the simulation as a function of halo mass.  The resulting stellar mass function is steepened in the mass range of ultra-faint dwarfs due to the large scatter in stellar mass at low halo masses, which places higher stellar mass into some low mass halos.  This luminosity/mass range is currently being probed by surveys like DES, which has discovered 16 new ultra-faint dwarfs in its first two years \citep{DES1, DES2}, and the HSC-SSP, which has already discovered one new dwarf \citep{Homma2017}.  Additional dwarfs should be discovered by LSST when it comes online, allowing the stellar mass function in this range to be probed observationally.

Most importantly, the large scatter in our simulated SMHM demonstrates that the commonly-adopted assumption of a monotonic relationship between stellar mass and halo mass that is adopted in abundance matching breaks down at low masses.  At the faintest end of the SMHM probed by our simulations, a galaxy cannot be assigned a unique halo mass based solely on its stellar mass or luminosity. In the regime M$_{star}$($z=0$) $\lesssim 10^6$ M$_{\odot}$ ($M_V \sim -8$), dark matter halo masses may vary by a factor of 100 at a fixed stellar mass.  Thus, abundance matching of galaxies below M$_{star} =$ 10$^6$ M$_{\odot}$ must consider the effect of large scatter.

The lack of a monotonic relation between stellar mass and halo mass has implications for interpreting observations of dwarf galaxies.  SMHM relations derived at higher masses cannot be extrapolated to infer the mass of a faint dwarf.  Instead, each faint galaxy must be regarded individually, using kinematics to infer halo mass.  Additionally, if a faint galaxy can be residing in a range of halo masses, it is no longer sufficient to assume a faint galaxy has been quenched by reionization. Rather, the halo mass in which the dwarf resides (or originally resided in, in the case of satellites) that matters.   By providing a model to stochastically populate the low mass halo function, we have opened the door to allow these issues to be studied more realistically.

\section*{Acknowledgements}
FDM acknowledges support from the Vanderbilt Initiative for Data-Intensive Astrophysics (VIDA) through a  
VIDA Postdoctoral Fellowship, from the University of Oklahoma, and from HST AR-13925 provided by NASA through a grant from the Space Telescope Science Institute, which is operated by the Association of Universities for Research in Astronomy, Incorporated, under NASA contract NAS5-26555. %AMB acknowledges support from NSF-AST-1411399. 
FG and and TRQ were partially supported by NSF award AST-1311956 and HST award AR-13264. FG was partially supported by NSF AST-1410012, HST AR-14281, and NASA NNX15AB17G. This research is part of the Blue Waters sustained-petascale computing project, which is supported by the National Science Foundation (awards OCI-0725070 and ACI-1238993) and the state of Illinois. Blue Waters is a joint effort of the University of Illinois at Urbana-Champaign and its National Center for Supercomputing Applications. This work is also part of a PRAC allocation support by the National Science Foundation (award number OCI-1144357). The authors thank Shea Garrison-Kimmel, Kelly Holley-Bockelmann, Michael Tremmel and Adi Zolotov for assistance and helpful comments. %, as well as the anonymous referee who will greatly improve this manuscript.

\bibliographystyle{mn2e}
\bibliography{bibref}

\end{document}